\documentclass[ApJL,twocolumn]{aastex62}

\usepackage{dcolumn, color, units, xspace, mathtools, physics, tensor, bm, lipsum, revsymb}
\usepackage[caption=false]{subfig}
\usepackage[normalem]{ulem}
\usepackage{import} 
\usepackage{multirow}
\usepackage{xspace}
\usepackage{acronym}
\usepackage{booktabs}

\usepackage{graphicx}	
\usepackage{amsmath}	

\begin{document}

\title[Noise models in GWB parameter estimation]{Choosing suitable noise models for nanohertz gravitational-wave astrophysics}

\author{Valentina Di Marco}
\affiliation{School of Physics and Astronomy, Monash University \\
Clayton VIC 3800, Australia}
\affiliation{OzGrav: The ARC Center of Excellence for Gravitational Wave Discovery \\
Clayton VIC 3800, Australia}
\affiliation{CSIRO, Space and Astronomy \\
PO Box 76, Epping, NSW 1710, Australia}

\author{Andrew Zic}
\affiliation{CSIRO, Space and Astronomy \\
PO Box 76, Epping, NSW 1710, Australia}
\affiliation{OzGrav: The ARC Center of Excellence for Gravitational Wave Discovery \\
Hawthorn VIC 3122, Australia}

\author{Ryan M. Shannon}
\affiliation{OzGrav: The ARC Center of Excellence for Gravitational Wave Discovery \\
Hawthorn VIC 3122, Australia}
\affiliation{Centre for Astrophysics and Supercomputing, Swinburne University of Technology \\
Hawthorn VIC 3122, Australia}

\author{Eric Thrane}
\affiliation{School of Physics and Astronomy, Monash University \\
Clayton VIC 3800, Australia}
\affiliation{OzGrav: The ARC Center of Excellence for Gravitational Wave Discovery \\
Clayton VIC 3800, Australia}

\author{Atharva D. Kulkarni}
\affiliation{OzGrav: The ARC Center of Excellence for Gravitational Wave Discovery \\
Hawthorn VIC 3122, Australia}
\affiliation{Centre for Astrophysics and Supercomputing, Swinburne University of Technology \\
Hawthorn VIC 3122, Australia}



\begin{abstract}
Accurately estimating the parameters of the nanohertz gravitational-wave background is essential for understanding its origin. 
The background is typically modeled with a power-law spectrum, parametrized with an amplitude $A$, which describes its intensity, and a spectral index $\gamma$, which describes how the background varies with frequency.
Different collaborations have produced varied estimates of $\gamma$, some in tension with the value of $\gamma = 13/3$ expected for circular, gravitational-wave-driven binary black holes.
However, estimates of $A$ and $\gamma$ can be affected by systematic errors and misspecified noise models.
We investigate how systematic errors---which may plausibly be present in pulsar-timing analyses---can shift inferences about $A, \gamma$. 
We demonstrate that conservatively incorporating noise sources into the model that are not actually present in the data does not produce bias inferences in practice. 
This addresses concerns that an overly complex noise model might lead to bias from a needlessly conservative prior.
Our results highlight the importance of using comprehensive noise models in pulsar timing analyses to ensure accurate and reliable parameter estimation of the gravitational-wave background.
\end{abstract}

 
\section{Introduction}

One of the objectives of pulsar timing experiments is to use extremely precise pulse arrival time measurements from millisecond pulsars to search for a gravitational-wave background in the nanohertz frequency band \citep{Foster_1990}. A promising source of this background is a population of inspiraling supermassive black hole binaries. \citep{Rajagopal_1995, Phinney_2001, Wyithe_2003}. 

Gravitational waves induce variations in the arrival times of pulses from the pulsars in an array. 
These variations are described by the timing residual cross-power spectral density $S_t(f)$, which is generally modeled as a power law $S_t(f) \propto f^{-\gamma}$, where $f$ represents the gravitational-wave frequency and $\gamma$ is the spectral index.
For a population of circular binary black holes whose orbital evolution is solely influenced by gravitational wave emission, we expect a spectral index consistent with $\gamma = 13/3$ \citep{Phinney_2001}.

Pulsar timing collaborations around the world have recently presented evidence for the angular correlations predicted by such a gravitational-wave background \citep{Antoniadis_2022, Agazie_2023_GW,  Reardon_2023_gw, Xu_2023,2025MNRAS.536.1489M}.

However, the spectral index values reported by various collaborations differ from the expected $\gamma \sim 13/3$. Specifically, the European Pulsar Timing Array shows a deviation of 1.7 $\sigma$ from this value, NANOGrav reports a 2.1 $\sigma$ deviation, and the Parkes Pulsar Timing Array shows a deviation of 1 $\sigma$.

These modest inconsistencies relative to $\gamma = 13/3$ could be explained with models of supermassive black hole evolution with stronger binary-environment interactions that could produce signals more consistent with current observations \citep{Ellis_2024}. Similarly, astrophysical explanations and alternative noise analysis methodologies have been explored to address inconsistencies in the detected signal amplitude \citep{Agazie_2023, Liepold_2024, Sato-Polito_2024, Goncharov_2024, SatoPolito_2025}.

Another possible justification for the discrepancy could be that the observed signal originates from more exotic sources other than supermassive black holes, such as cosmic strings and phase transitions in the early Universe. 
In \cite{Afzal_2023} the NANOGrav collaboration has explored the possibility that the observed gravitational-wave background could also be explained by cosmological phenomena, such as cosmic inflation \citep{Vagnozzi_2020, Kuroyanagi_2015, Benetti_2022}, scalar-induced gravitational waves \citep{Vaskonen_2020, DeLuca_2021, Kazunori_2021}, cosmological phase transitions \citep{Witten_1984, Hogan_1986}, cosmic strings \citep{Damour_2000, Damour_2001, Sanidas_2013}, or domain walls \citep{Vilenkin_1981, Preskill_1991, Kawasaki_2011}. 
However, they also caution that the observed tension could be resolved with more accurate modeling and additional data. 

In \cite{EPTA_2024_new_physics} the European Pulsar Timing Array (EPTA) and Indian Pulsar Timing Array (InPTA) collaborations investigated potential cosmological origins for the gravitational-wave background, including cosmic strings, first-order phase transitions, primordial inflation, ultralight dark matter, as well as a population of inspiralling supermassive black hole binaries (SMBHBs). Their findings indicate that the signal is best explained as originating from binary black holes.  
This is because most alternative signals predict backgrounds steeper than those from SMBHBs.

In pulsar timing, detecting a gravitational-wave signal requires the development of comprehensive noise models that can identify and account for other noise sources contributing to timing variations \citep{Lentati_2016, Goncharov_2021B, Agazie_2023_noise, EPTA_pulsar_noise, Reardon_2023_null_hyp, Miles_2024_noise}.
However, pulsar timing array data are complex, and the noise processes involved are not fully understood. 
Correctly characterizing the parameters of the known noise process is difficult, and there may also be unidentified noise sources in the data.
Different collaborations use models with different degrees of complexity.

It is already known that misspecified noise models have an adverse effect on the detection of a common spectrum. For example, \cite{Hazboun_2020} demonstrated that the specifics of red noise modeling, including amplitude priors and the selection of pulsars with red noise, significantly influence gravitational-wave background statistics, affecting both credible intervals and amplitude estimates of the gravitational-wave background. \cite{Goncharov_2021a} and \cite{Zic_2022} both demonstrated, through simulation, that search methods can falsely detect a common red process in pulsar timing array datasets where it is absent. It has also been shown that noise misspecifications such as unaccounted for chromatic noise and low-amplitude instrumental jumps, affect the optimal statistic used for detecting correlated signals \citep{Di_Marco_2024}.
This can lead to overly conservative detection confidence \citep{Di_Marco_2024}.

There are a number of methods to assess the validity of noise models. Recent work \citep{Meyers_2023, Vallisneri_2023} highlights the use of posterior predictive checks to evaluate the predictive performance of models in pulsar timing arrays, examining components such as the spectral shape, correlation patterns, and timing residuals to identify potential misspecifications. These are important methods to improve the astrophysical inferences as well as the reliability of detection claims.

The critical impact on detection outcomes due to the degeneracies between different noise sources, specifically, dispersion measure noise and the achromatic gravitational-wave background was also recently addressed \citep{Ferranti_2024}. Importantly, the degeneracy was linked to a loss of statistical significance for the recovered gravitational-wave signal and was shown to bias the gravitational-wave background parameters, favoring a higher and flatter spectral shape.

In this paper, we investigate how the choice of noise models in pulsar timing analysis affects the spectral properties of the common-spectrum process. Specifically, we examine whether the observed tension in posterior estimates could be explained by misspecification of the noise model.
We simulate a set of pulsars with various noise sources and a gravitational-wave background, and then estimate the parameters of the gravitational-wave background with a plausibly misspecified noise model.
We compare this result with the one obtained using a correctly specified noise model.

The remainder of this paper is organized as follows: in Section \ref{Simulations}, we describe the procedure for simulating a dataset and the model used to estimate the posterior parameters of the gravitational-wave background.
In Section \ref{Results} we present our results showing how misspecified models have an effect on those posterior estimation.
Concluding remarks are provided in Section \ref{Conclusions}.

\section{Simulations}\label{Simulations}
We simulate datasets reflective of the PPTA DR3 analysis, incorporating several well-known stochastic signals to illustrate their influence on gravitational-wave parameter estimation. For simplicity, we do not aim to explore all potential noise sources exhaustively. The simulations included intrinsic red noise from pulsars, dispersion measure variations, chromatic noise, solar wind fluctuations, and instrumental jumps.  We describe these noise processes in more detail below.

We start with a set \texttt{Tempo2}-format \citep{Hobbs_2006}  pulsar ephemerides, and simulate times of arrival for 30 pulsars in the PPTA array with white noise and a gravitational-wave signal with $\log_{10} A_{\text{GW}} = -14.70$ and spectral index $\gamma_{\text{GW}} = 13/3$, using the package \texttt{PTAsimulate}.\footnote{https://bitbucket.org/psrsoft/ptasimulate} 
For each pulsar, we then inject intrinsic red noise and dispersion-measure noise in order to work with a more realistic dataset that broadly reproduce the dataset in the PPTA gravitational wave analysis \citep{Reardon_2023_gw}.

Pulsar-intrinsic red noise, also known as timing noise or spin noise, is thought to be associated with stochastic irregularities in pulsar rotation and is often modeled as a stationary stochastic signal. This is a temporally correlated noise that is the same at all radio frequencies, possibly due to interactions between the neutron star's crust and its superfluid core \citep{Jones_1990}. 
Spin noise in millisecond pulsars is generally well modeled by a power-law process \citep{Goncharov_2020}. 
However, some pulsars show quasiperiodic behavior or spectral turnovers, suggesting that additional modeling could be needed \citep{Lyne_2010, Parthasarathy_2019}.

Dispersion measure variations \citep{Keith_2013} are caused by changes in the column density of free electrons along the line of sight between the pulsar and Earth. As radio waves pass through the ionized interstellar medium, higher-frequency components travel faster than lower-frequency ones, leading to frequency-dependent delays in the pulse arrival times. This time delay $\Delta t$ is directly proportional to the column density but inversely proportional to the square of the radio frequency $\nu$, so that $\Delta t \propto \nu^{-2}$ and is also modeled as a power law process.
We inject noise in the \texttt{PTAsimulate} simulations using the \texttt{libstempo}\footnote{https://github.com/vallis/libstempo} package. 
The power-law parameters we used are listed in Table \ref{table:noise}. These are the same parameters as those estimated in \cite{Reardon_2023_null_hyp}, which can be found in Table~1 of that paper.

\begin{table}
\centering
\begin{tabular}{lcccc}
\toprule
\textbf{Pulsar} & \(\log_{10} A_{\text{TN}}\) & \(\gamma_{\text{TN}}\) & \(\log_{10} A_{\text{DM}}\) & \(\gamma_{\text{DM}}\) \\
\midrule
J0030+0451	&	-16.3	&	3.4	&	-16.8	&	3.4	\\
J0125-2327	&	-16.9	&	3.3	&	-13.4	&	3.2	\\
J0437-4715	&	-14.3	&	3.3	&	-13.48	&	2.5	\\
J0613-0200	&	-15.4	&	5.9	&	-13.6	&	2.4	\\
J0614-3329	&	-16.4	&	3.2	&	-13.8	&	5.0	\\
J0711-6830	&	-13.1	&	1.2	&	-14.1	&	3.2	\\
J0900-3144	&	-16.5	&	3.4	&	-12.7	&	2.0	\\
J1017-7156	&	-16.1	&	3.2	&	-12.89	&	2.3	\\
J1022+1001	&	-16.6	&	3.2	&	-13.8	&	2.5	\\
J1024-0719	&	-17.1	&	3.1	&	-13.9	&	3.5	\\
J1045-4509	&	-14.5	&	1.4	&	-12.38	&	2.9	\\
J1125-6014	&	-14.2	&	3.9	&	-13.2	&	3.6	\\
J1446-4701	&	-17.0	&	3.1	&	-16.9	&	2.7	\\
J1545-4550	&	-16.6	&	3.3	&	-13.7	&	4.4	\\
J1600-3053	&	-15.9	&	3.4	&	-13.2	&	2.3	\\
J1603-7202	&	-17.2	&	2.9	&	-13.2	&	2.3	\\
J1643-1224	&	-12.7	&	0.6	&	-12.9	&	2.3	\\
J1713+0747	&	-17.5	&	2.9	&	-13.9	&	0.3	\\
J1730-2304	&	-16.0	&	2.3	&	-13.5	&	2.5	\\
J1744-1134	&	-15.9	&	2.3	&	-14.2	&	3.2	\\
J1832-0836	&	-16.9	&	3.2	&	-13.5	&	4.5	\\
J1857+0943	&	-14.7	&	4.9	&	-13.3	&	2.4	\\
J1902-5105	&	-16.12	&	3.4	&	-13.1	&	1.4	\\
J1909-3744	&	-14.7	&	4.0	&	-13.66	&	2.0	\\
J1933-6211	&	-16.1	&	3.4	&	-16.9	&	3.3	\\
J1939+2134	&	-14.6	&	6.2	&	-12.91	&	2.8	\\
J2124-3358	&	-14.9	&	4.7	&	-17.3	&	3.0	\\
J2129-5721	&	-16.6	&	3.4	&	-13.7	&	3.1	\\
J2145-0750	&	-14.5	&	4.2	&	-13.5	&	1.8	\\
J2241-5236	&	-14.5	&	3.0	&	-14.0	&	2.7	\\
\bottomrule
\end{tabular}
\caption{Values of the injected parameters for timing noise and dispersion measure.}
\label{table:noise}
\end{table}

In addition to dispersion-measure noise, other sources of chromatic red noise have been identified, with frequency dependencies described by a chromatic index which may differ from the typical dependence of dispersion measure variations. 
Various astrophysical mechanisms, such as scattering in the interstellar medium, can introduce chromatic red noise with steeper frequency dependencies, ranging from $\nu^{-4}$ to $\nu^{-6.4}$ \citep{Shannon_2017}.
Unmodeled variations in the solar wind can also manifest as excess dispersion measure, which can affect the pulsar times of arrival.
Finally, instrumental-origin jumps may be present in the datasets, such as those arising from changes to the telescope's back-end instruments \citep{Manchester_2013, Arzoumanian_2018}. 
While high-amplitude jumps are easily detectable, smaller amplitude jumps may go undetected, presenting a challenge when accurately accounting for all timing offsets.

We then inject other noise sources, specifically chromatic noise \citep{Lentati_2016, Cordes_2010, Shannon_2017, Goncharov_2021a}, stochastic variations in  the solar wind \citep{You_2007, Tiburzi_2021, Hazboun_2022, Nitu_2024}, and instrumental jumps \citep{Manchester_2013, Arzoumanian_2015, Kerr_2020, Sarkissian_2011} to examine the effects of noise misspecification. These noise sources are not always incorporated into the posterior analysis.
We evaluate how the posteriors for $A, \gamma$ are affected when these optional noise sources are present in the data, but are left out of the analysis.

We model chromatic noise in all pulsars as a power law with amplitude $\log_{10} A_{\text{CH}} = -13.5$ and $\gamma_{\text{CH}} = 2.5$ and chromatic index, (the spectral index for the radio spectrum), of $4$.
Solar wind variations are also modeled as a power law with parameters $\log_{10}A_{\text{SW}} = -6$ and $\gamma_{\text{SW}}= 1.6$ using the model presented in \citet{Hazboun_2022}.
The jumps are sampled from a uniform distribution over the range [0, $\unit[200]{ns}$]. 
Their arrival times follow a Poisson distribution with an expected occurrence rate of $\unit[10]{yr^{-1}}$.
Since the jumps originate from instrumental sources, we assume their arrival times and amplitudes to be identical for all pulsars in the array and to occur exclusively at the $\unit[3100]{MHz}$ radio frequency, simulating the effects previously observed in single receiving systems of Parkes \citep{Manchester_2013, Arzoumanian_2015, Kerr_2020}.

We carry out Bayesian parameter estimation with the \texttt{enterprise}\footnote{https://github.com/nanograv/enterprise} package. 
In the misspecified model, we search only for white noise, timing noise, and dispersion measure variations. In the correctly specified model, we include all the injected noise sources.
Since the common red noise provides most of the information about the spectral properties of the gravitational-wave background in full Hellings-Downs PTA searches, we estimate the gravitational-wave parameters using this uncorrelated common signal instead of the correlated one \citep{Hourihane_2023}. It is worth mentioning that the correlations analysis could provide valuable insights into identifying and safeguarding against noise misspecification. These correlations represent a robust assumption within the model, one that noise is unlikely to satisfy. In contrast, a common uncorrelated signal is more easily generated by noise.

\section{Results}\label{Results}
\begin{figure}
    \centering
\includegraphics[clip,width=\columnwidth]{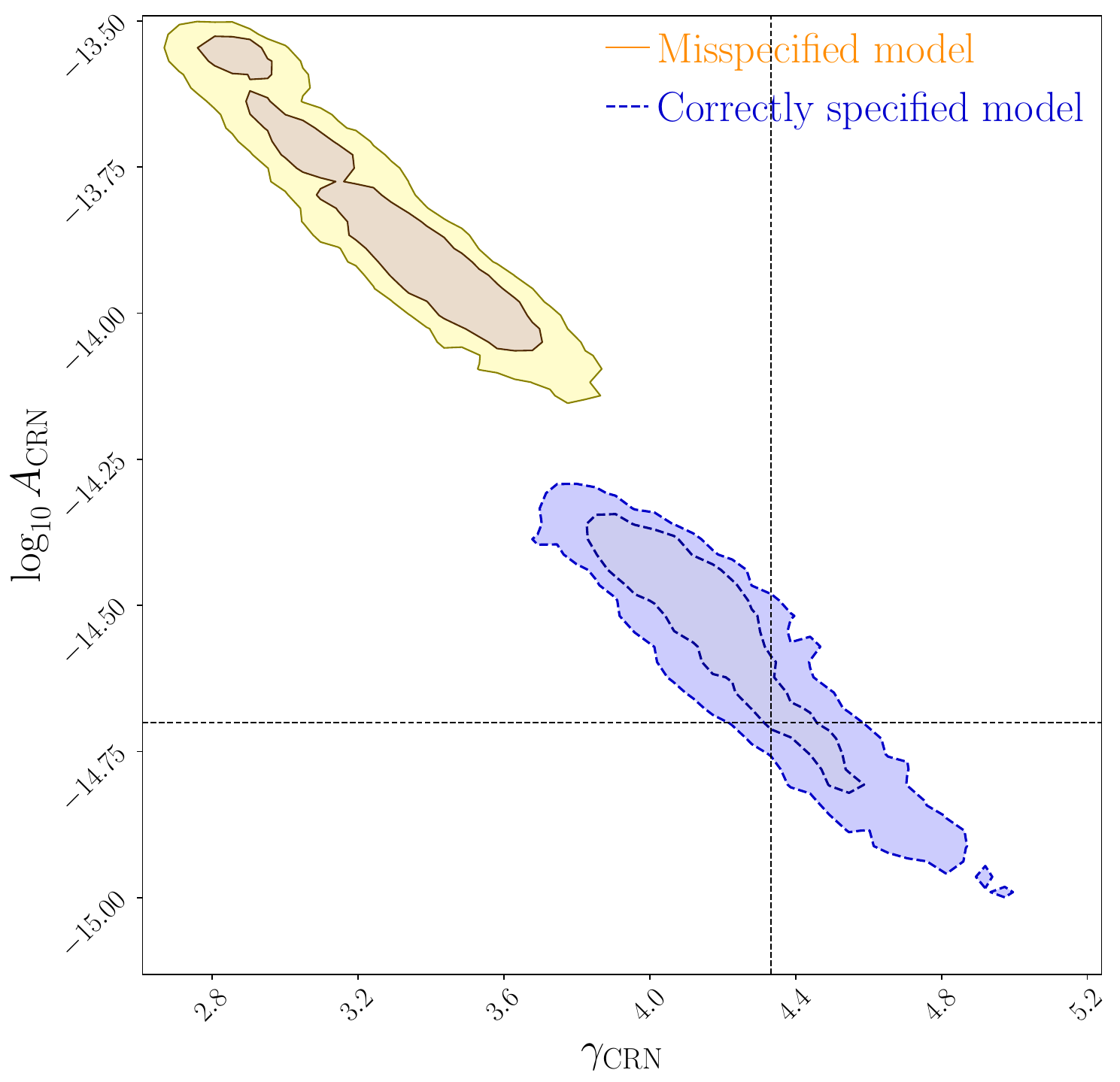}
    \caption{
    Credible intervals for $\log_{10} A, \gamma$ for a simulation containing chromatic noise in all pulsars.
    The solid curves show the results obtained with a misspecified noise model (not including chromatic noise) while the solid curves show the results for a correctly specified model.
    The true values of the gravitational wave signal are marked by the horizontal and vertical dashed lines.}
    \label{fig:ch_all}
\end{figure}

Figure~\ref{fig:ch_all} shows the posterior distribution for the spectral index (horizontal axis) and the log amplitude (vertical axis) of the gravitational-wave background. The solid contours represent the misspecified model, and the dashed contours correspond to the correctly specified model, with the injected values indicated by a dashed line. 
When chromatic noise is simulated across all pulsars, but not accounted for in the \texttt{Enterprise} models, the spectral index is significantly underestimated and the amplitude of the gravitational-wave signal is overestimated.
The misspecified credible interval does not include the true values of $\log_{10} A, \gamma$.

\begin{figure}
    \centering
\includegraphics[clip,width=\columnwidth]{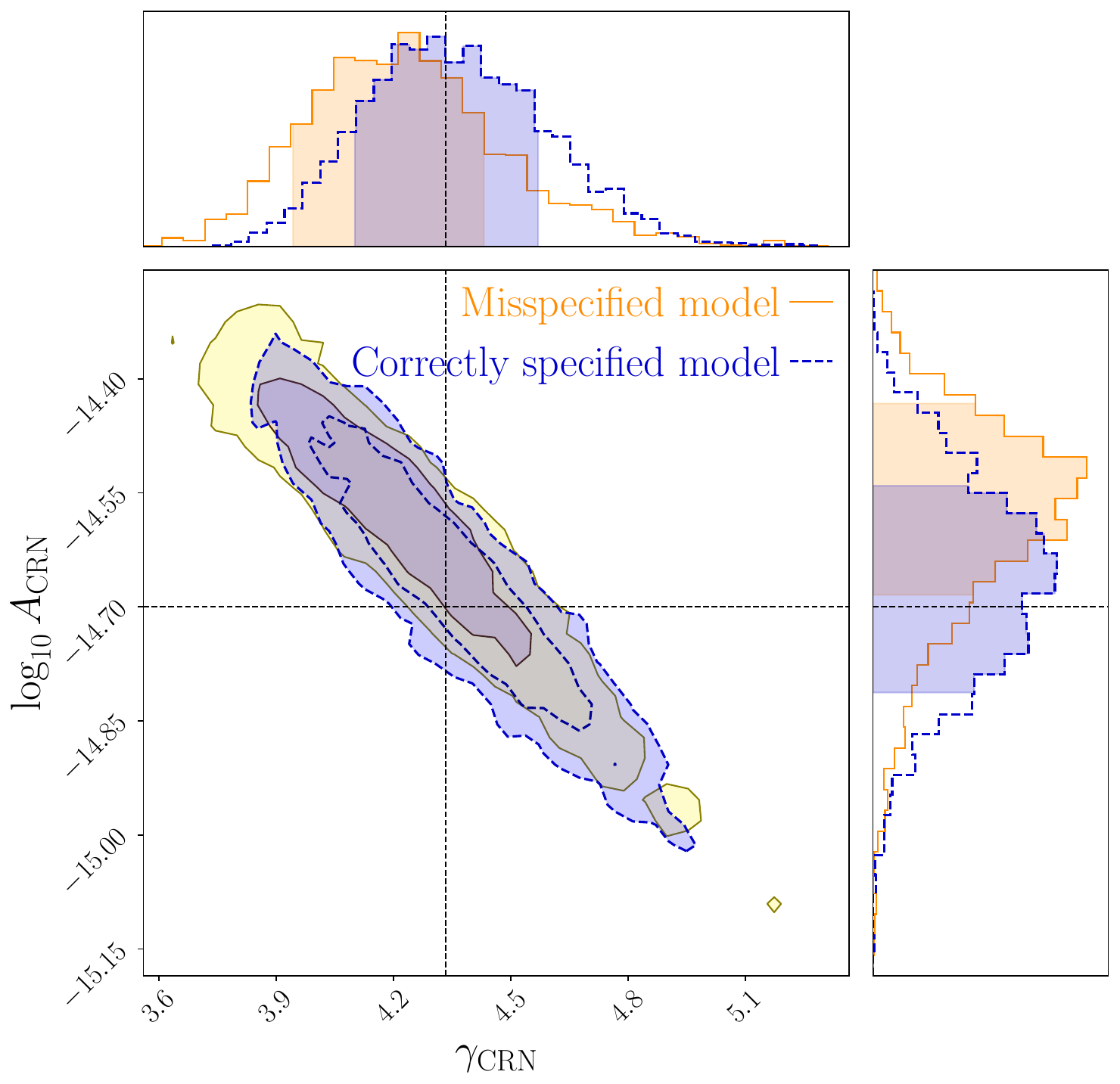}
    \caption{
    Credible intervals for $\log_{10} A, \gamma$ for a simulation containing jumps in all pulsars.
    The solid curves show the results obtained with a misspecified noise model (not including jump noise) while the solid curves show the results for a correctly specified model.
    The true values of the gravitational wave signal are marked by the horizontal and vertical dashed lines.
    }
    \label{fig:jumps}
\end{figure}

We observe similar results for instrumental jumps; see Fig.~\ref{fig:jumps}. 
While the effect of the jumps with the simulated frequency and amplitude is less pronounced than the chromatic noise, the presence of unmodeled jumps results in a shallower spectrum and an overestimated amplitude.

When evaluating the effect of an unmodelled solar wind, we find no significant difference between the results from the misspecified and correctly specified models. Evidently, the misspecified noise caused by the solar wind gets absorbed into the dispersion measure variations.

\begin{figure}
    \centering
\includegraphics[clip,width=\columnwidth]{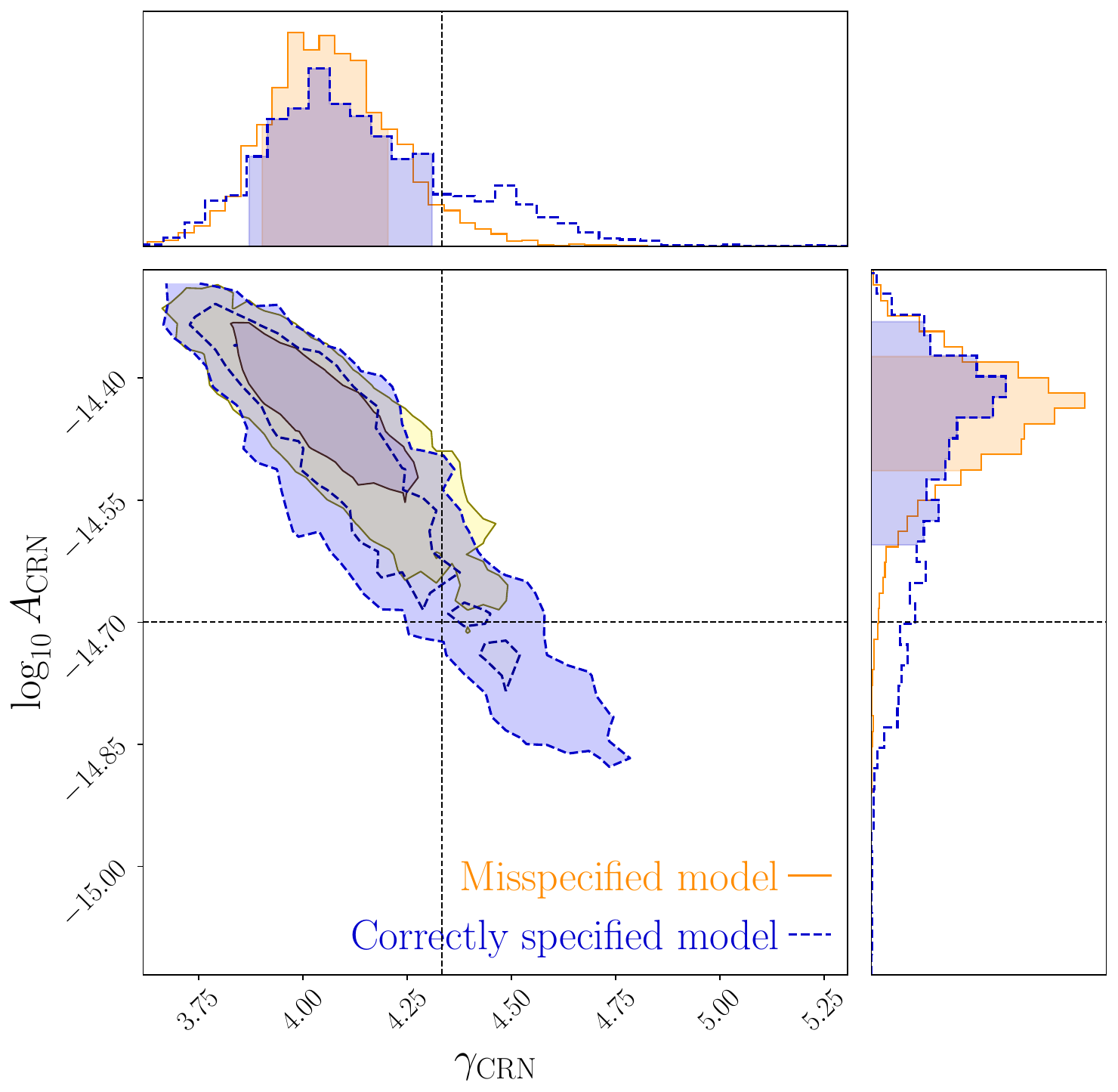}
    \caption{
    Credible intervals for $\log_{10} A, \gamma$ for a simulation containing solar wind in all pulsars.
    The solid curves show the results obtained with a misspecified noise model (not including solar wind) while the solid curves show the results for a correctly specified model.}
    \label{fig:solar}
\end{figure}

\begin{figure}
    \centering
\includegraphics[clip,width=\columnwidth]{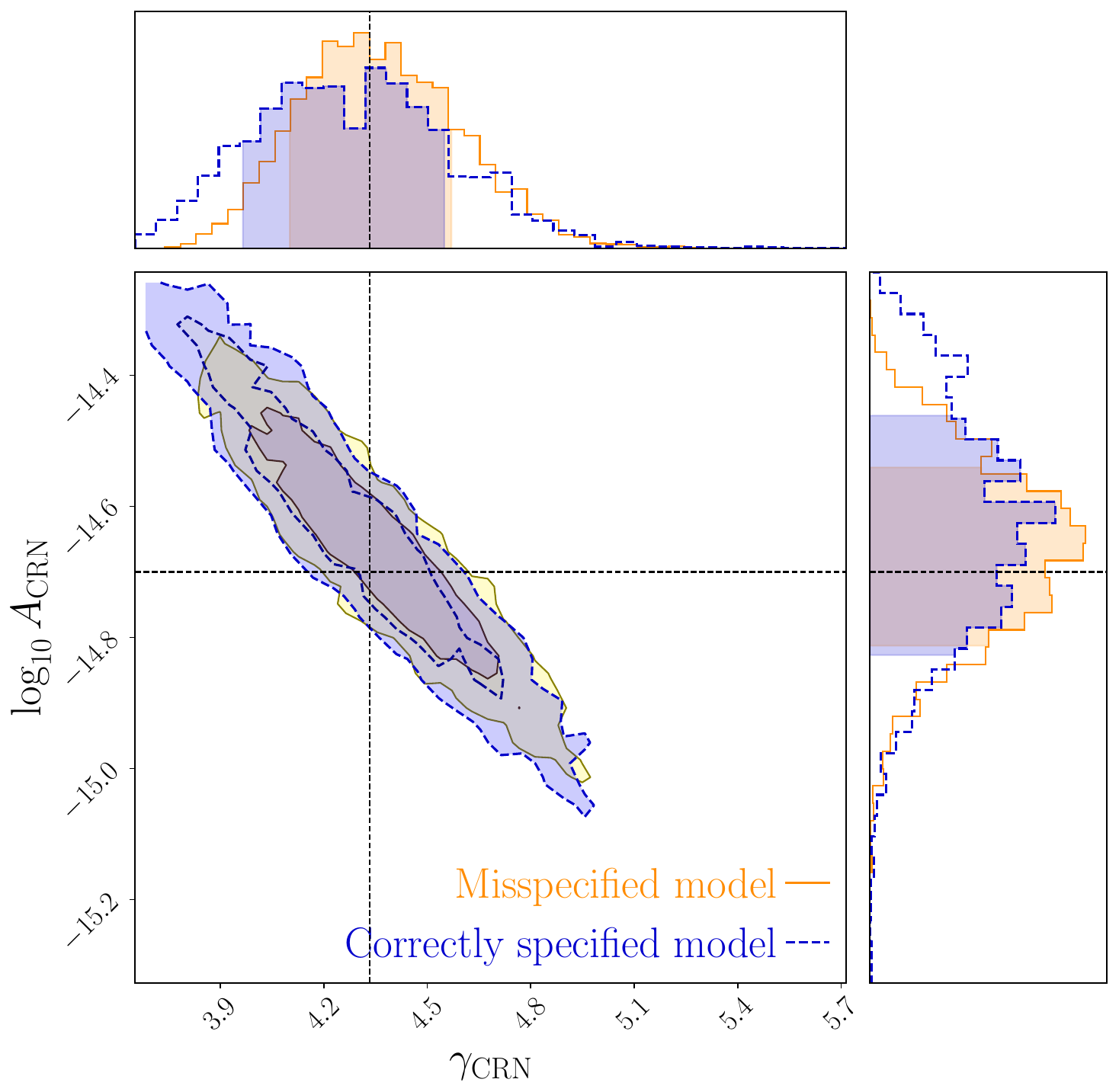}
    \caption{
    Credible intervals for $\log_{10} A, \gamma$ for a simulation containing white noise, red noise, and dispersion measure.
    The solid curves show the results obtained with a correctly specified noise model while the solid curves show the results for a misspecified model containing extra sources of noise (chromatic noise and solar wind).
    The true values of the gravitational wave signal are marked by the horizontal and vertical dashed lines.
    }
    \label{fig:complex}
\end{figure}

\begin{figure}
    \centering
\includegraphics[clip,width=\columnwidth]{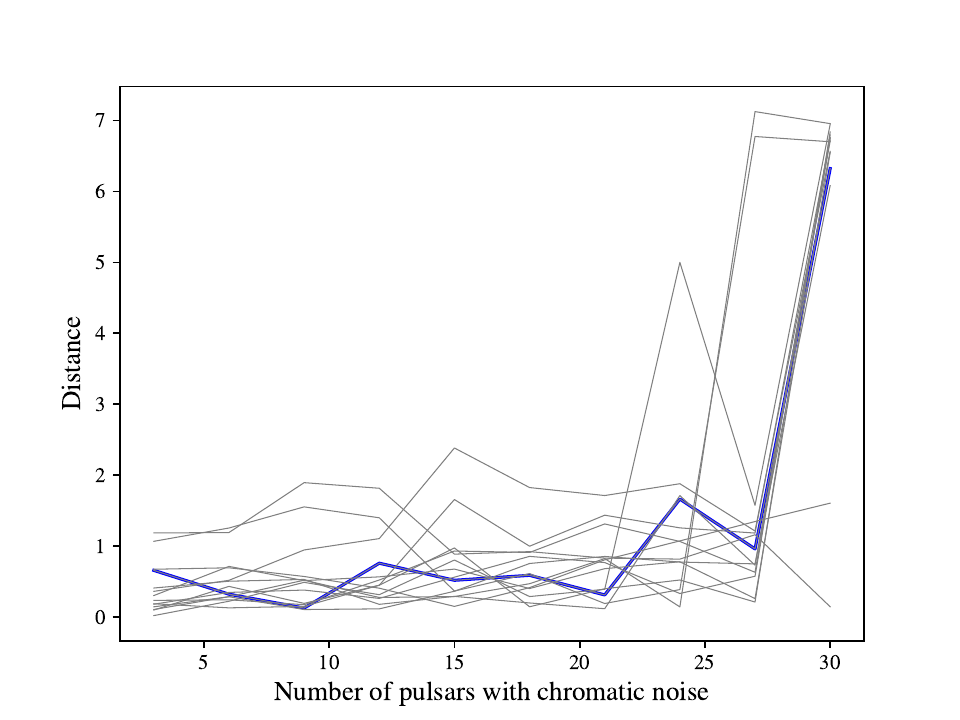}
    \caption{
    Mahalanobis distance between the marginal distributions of $\log_{10}A$, $\gamma$ for correctly specified and misspecified models as a function of the number of pulsars affected by misspecification. Each point represents the mean and standard deviation of a distribution of Mahalanobis distances, calculated between the correctly specified model's distribution and the points of a misspecified model. The comparisons are repeated across multiple pulsar orderings, shown in grey.
}
    \label{fig:distance}
\end{figure}

In the second part of our analysis, we investigate the effects of incorporating noise sources into the parameter estimation models that are not present in the simulated data. 
In other words, we aim to determine whether a conservative model that accounts for non-existent noise sources in the models introduces bias in the parameter estimation process. 
To do so, we carry out a simulation containing only intrinsic red noise and dispersion measure variations but analyze it with a model that also includes chromatic noise and solar wind variations in all pulsars.
As can be seen in Figure \ref{fig:complex}, the posterior distributions obtained from the two models show no significant differences, suggesting that the inclusion of unnecessary noise sources does not necessarily affect the accuracy parameter estimation.
In the final part of our analysis, we aim to determine the level of misspecification required to induce a significant shift in the posterior estimates of the gravitational-wave signal. 
To investigate this, we inject chromatic noise into groups of three pulsars at a time, progressively increasing the number of affected pulsars until all pulsars have chromatic noise in their simulated times of arrival. 
For each simulation, we compute posterior estimates for the background using both a misspecified model that did not include chromatic noise, and a correctly specified model that accounts for the presence of chromatic noise. 
We then compute the Mahalanobis distances \citep{Mahalanobis_1936} between the posterior distribution of the correctly specified model and that of the misspecified model. Following the approach used by the IPTA \citep{IPTA_GW}, we implement the Mahalanobis distance $D_M$ as a generalized measure of separation between two 2D distributions, defined as
\begin{equation}
    D_M = \sqrt{(\vec{\mu_1} - \vec{\mu_2})^{T} \Sigma^{-1} (\vec{\mu_1} - \vec{\mu_2})}
\end{equation}
were $\vec{\mu_1}$ and $\vec{\mu_2}$ are the mean vectors of the distributions we compare, and $\Sigma = (\Sigma_1 + \Sigma_2)/2$ is the joint covariance.
This ensures that the distance accounts for correlations between variables while normalizing for differences in scale.
We then plot these distances against the number of pulsars affected by misspecification and repeat the analysis using various pulsar orderings to confirm that no single pulsar disproportionately influenced the results (Figure \ref{fig:distance}).
Our findings reveal that the distance does not increase linearly with the number of pulsars affected by chromatic noise. 
Instead, an increase in distance occurs when the number of affected pulsars approaches 27. 
We currently lack an explanation for the fact that this happens when this ``misspecification threshold '' of 27 pulsars is reached.

\section{Conclusions}\label{Conclusions}
In this study, we examine how noise model misspecification impacts the estimation of gravitational-wave background parameters in pulsar timing array (PTA) data. 
Our results demonstrate that incorrect or incomplete noise models can lead to significant bias in the spectral index and amplitude of the gravitational-wave signal. 
Specifically, if one analyzed  PPTA DR3, but ommitted chromatic noise, instrumental jumps, or solar wind effects, one would likely underestimate the spectral index and overestimate the amplitude. 
Such misspecification may plausibly exist in PTA analyses. 
This investigation highlights the potential risks of neglecting more complex noise in our models.
Analyses that fail to incorporate adequate noise models may produce biased estimates of gravitational-wave parameters.
Our findings align with \citet{Reardon_2023_null_hyp}, which showed that the use of overly simplified models can significantly impacts the recovered spectral characteristics compared to a more comprehensive noise model.

Our results suggest that recent claims of spectral indices that deviate from the canonical value of $\gamma = 13/3$ might be more indicative of inadequacies in noise modelling rather than of new physics. 
Differences in noise modeling approaches between PTAs may also contribute to tension in results obtained by different collaborations. 
A recent study from the IPTA \citep{Agazie_2024} highlights the benefits of adopting a standardized noise model across all pulsars and PTAs, showing that this approach reduces tensions in pulsar noise parameters and improves the consistency of gravitational wave parameter estimates. 
However, the standardized model used in that study did not include all possible noise sources. 
It accounted only for intrinsic pulsar red noise, interstellar dispersion measure variations, a deterministic solar wind model, and a fixed spectral index of common uncorrelated red noise. 

We also show that incorporating noise sources absent from the data in our models has a minimal impact on parameter estimation. Therefore, using more comprehensive noise models in the analysis likely poses little risk.
Furthermore, we find evidence of a "misspecification threshold," where the number of pulsars affected by unmodeled noise processes must exceed a certain level before significantly affecting the posterior estimates. 
This suggests that modest noise misspecifications may not significantly impact the results, but there is a critical point at which such errors become detrimental to the analysis. 
We speculate that in large datasets with a high number of pulsars, the risk of reaching this threshold could be higher, as small misspecifications in individual pulsars could accumulate and collectively push the analysis beyond the threshold.

While we model several known noise processes, such as chromatic noise, solar wind, and instrumental effects, the complex nature of pulsar timing arrays suggests that additional, unidentified noise sources could exist. These unknown factors could also add up to introduce biases in the posterior estimates of the gravitational wave parameter.
A related concern is the potential misspecification introduced by the lack of hierarchical modeling in PTA analyses \citep{Van_Haasteren_2024_A}.
Hierarchical Bayesian modeling offers a framework to jointly describe ensemble and individual pulsar properties, mitigating these biases. Incorporating these hierarchical models in a comprehensive noise model analysis could further improve the robustness of gravitational wave parameter estimation.

Building on the concept of the ``misspecification threshold'' introduced in that paper, it is important to highlight a recent study on noise analysis in pulsar timing \citep{Van_Haasteren_2024_B}. That study raises concerns about ``circular analysis'' in pulsar timing, where posterior distributions from single-pulsar noise analyses are used as priors for the array-wide analysis. This approach is sometimes employed to streamline the computationally demanding process of analyzing the full pulsar array by eliminating unnecessary model components. However, this practice can lead to issues; for example, a red noise component might be excluded in some pulsars if the Bayes factor used is just below the chosen threshold, even though intrinsic red noise may still exist. It would be interesting to investigate the extent of this issue and determining the minimum number of affected pulsars required to induce model misspecifications in this case.

\section*{Acknowledgements}
We acknowledge and pay respects to the Elders and Traditional Owners of the land on which this work has been performed, the Bunurong, Wadawurrong and
Wurundjeri People of the Kulin Nation and the Wallumedegal People of the Darug Nation. The authors thank Patrick Meyers for his valuable suggestions and for recommending additional citations and clarifications that greatly improved the manuscript.
The authors are supported via the Australian Research Council (ARC) Centre of Excellence CE170100004 and CE230100016.
E.T. is additionally supported through ARC DP230103088.
V.D.M. receives support from the Australian Government Research Training Program. AK acknowledge support through ARC grant
CE170100004.
This work was performed on the OzSTAR national facility at Swinburne University of Technology. The OzSTAR program receives funding in part from the Astronomy National Collaborative Research Infrastructure Strategy (NCRIS) allocation provided by the Australian Government.

\section*{Data Availability}

No data were used in the analysis. Simulated data and code used in this analysis will be made available upon reasonable request.


\bibliography{refs}
\bibliographystyle{aasjournal}



\end{document}